\documentclass[a4wide]{article}
\usepackage{amssymb,epsfig}

\title{\bf Physical Diffeomorphisms in Loop Quantum Gravity}
\author{Tim A. Koslowski
        \\ Institut f\"ur Theoretische Physik und Astrophysik
        \\ Universit\"at W\"urzburg 
        \\ 97074 W\"urzburg, Germany
        \\ {\texttt {tim@physik.uni-wuerzburg.de}} }
\date{ \today}

\newtheorem{defi}{Definition}
\newtheorem{lem}{Lemma}
\newtheorem{cor}{Corollary}

\begin{document}

\maketitle

\thispagestyle{empty}

\subsubsection*{Abstract:} We investigate the action of diffeomorphisms in the context of Hamiltonian Gravity. By considering how the diffeomorphism-invariant Hilbert space of Loop Quantum Gravity should be constructed, we formulate a physical principle by demanding, that the gauge-invariant Hilbert space is a completion of gauge- (i.e. diffeomorphism-)orbits of the classical (configuration) variables, explaining which extensions of the group of diffeomorphisms must be implemented in the quantum theory. It turns out, that these are at least a subgroup of the stratified analytic diffeomorphisms. Factoring these stratified diffeomorphisms out, we obtain that the orbits of graphs under this group are just labelled by their knot classes, which in turn form a countable set. Thus, using a physical argument, we construct a separable Hilbert-space  for diffeomorphism invariant Loop Quantum Gravity, that has a spin-knot basis, which is labelled by a countable set consisting of the combination of knot-classes and spin quantum numbers. It is important to notice, that this set of diffeomorphism leaves the set of piecewise analytic edges invariant, which ensures, that one can construct flux-operators and the associated Weyl-operators. A note on the implications for the treatment of the Gauss- and the Hamilton-constraint of Loop Quantum Gravity concludes our discussion.

\newpage
% \section{Introduction}
  % Introduction

\section{Introduction}

Loop Quantum Gravity (see e.g. \cite{rovelli}\cite{thiemann}\cite{ash-lew}) has been developed as a quantization of General Relativity on the basis of mathematical consistency. A key ingredient is the minimalistic ansatz, that takes canonical General Relativity in the connection formulation chooses a reasonable set of basic variables and then studies the Hilbert-space representations of the resulting Poisson- resp. Weyl-algebra.
\\
A very strong constraint for the construction of the Hilbert-space, which in fact makes the representation unique (up to technical assumptions), is the covariance under diffeomorphisms. It turns out, that the kinematical Hilbert-space (before gauge fixing) is non-separable. The hope is, that it becomes separable after the diffeomorphisms are factored out. This result is however disputed in literature: On the one hand Fairbairn and Rovelli\cite{fair-rov} constructed a Hilbert-space of diffeomorphism orbits, which is indeed separable. The group of so called ''extended diffeomorphisms'' ad hoc postulated to be the group of quantum diffeomorphisms, and the physical argument comes a posteriori by showing, that there exists a version of quantum geometry, that transforms covariantly under the extended diffeomorphisms.
\\
Ashtekar and Lewandowski\cite{ash-lew} (and others) on the other hand argued, that there is no physical reason to extend the group of diffeomorphisms. A primary reason is, that a quantum connection can be reconstructed from the holonomies along piecewise analytical edges, which turns out to be useful in order to define the flux-operators. This category is preserved under analytical diffeomorphisms, thus the restriction. (In this paper, we will mostly not refer to any differentiability category, since our construction can already be done in the category of analytical objects. Thus, we will usually use the term smooth and mean any differentiability class and analytical.) An a posteriori argument is, that there exists an apparently physical version of quantum geometry, that does not transform covariantly under the extended diffeomorphisms. 
The main argument is: Diffeomorphisms act as linear transformations on the tangent space, thus a  can only map a triple of tangent vectors on another triple. However by considering spin networks, that have vertices with valence larger than three, one can construct invariants of the tangent vectors of the adjacent edges, that are invariant under linear transformations.
This dispute and the resulting non-separability of the Hilbert-space of Loop Quantum Gravity has led to criticism of Loop Quantum Gravity as a whole. 
\\
In this paper, we try to give an answer to the question, which group of diffeomorphisms should be used in the quantum theory. Our argument is entirely physical: A Quantum Gauge theory should be built from completions of series of classical gauge orbits. Thus, a physical gauge group should factor out all those limits, that do not arise as completions of gauge orbits. The diffeomorphisms are pure gauge in General Relativity, and hence we have to treat them in this way.
\\
Our investigation shows, that a subgroup of the stratified analytic diffeomorphisms leave the classical theory invariant and can be extended to be the group of gauge transformations, that has the physical property described above. Thus, we use this group of diffeomorphisms as the group of gauge transformations in the quantum theory. It turns out, that the orbits of graphs are indeed the knot-classes of the graphs, which form a countable set. Thus, we conclude similar to Fairbairn and Rovelli \cite{fair-rov}, that the diffeomorphism invariant Hilbert-space of Loop Quantum Gravity is separable. Our case is closed by a quick observation, that there even exists a version of quantum geometry, that transforms covariantly under stratified diffeomorphisms. Thus, we argue, that there is a separable Hilbert-space for Loop Quantum Gravity, that is constructed entirely from physical principles.
% \section{Preliminaries}
  % Preliminaries

\section{Preliminaries}

In this section, we will introduce the notions used in this paper and expose the question, that we will try to answer.

\subsection{Classical Ashtekar Gravity}

Let $(\mathbb M,g)$ be a 4-dimensional globally hyperbolic Riemannian manifold, such that it foliates into spatial slices $(\Sigma,q)$. Then, we can split the 4-dimensional metric $g$ into:
\begin{equation}
  g=(q_{ab}N^a N^b -N^2) dt\otimes dt + q_{ab} N^a dt \vee dx^a + q_{ab} dx^a \otimes dx^b,
\end{equation}
where $q$ denotes the spatial metric, $N$ the lapse function and $N^a$ the shift vector field. Thus, we have the second fundamental form $K$:
\begin{equation}
  K_{ab}=\frac{\dot{q}_{ab} -\mathcal{L}_{\vec{N}} q_{ab} }{2N}.
\end{equation}
Writing the spatial metric in terms of dreibeins $e: q_{ab}=\delta_{ij} e^i_a e^j_b$, lets us define the spin connection $\Gamma^i_a$ through: $\partial_a e^j_b +\epsilon_{ijk} \Gamma^k_a e^j_b-\Gamma^c_{ab}e^j_c$. Then it can be shown, that an $SU(2)$-gauge-field $A$ and a densitized inverse triad furnish a canonical system of variables (for General Relativity), and that their relation to the geometrical quantities are:
\begin{equation}
  A^j_a=\Gamma^j_a +\beta K_{ab} e^b_j
\end{equation}
and
\begin{equation}
  E^a_i = \beta^{-1} \sqrt{det(q)}e^a_i.
\end{equation}
Let us smear these with a vector density $F^a_i$with values in $su(2)$ and with a covector test function $f^i_a$ with values in $\overline{su(2)}$:
\begin{equation}
  F(A):=\int_\Sigma d^3x F^a_i A^i_a \textrm{ and: } E(f):=\int_\Sigma d^3x E^a_i f^i_a.
\end{equation}
The Poisson structure for General Relativity $\{.,.\}$ is then equivalent to the definition:
\begin{equation}
  \begin{array}{rl}
    \{ E(f_1),E(f_2) \} = & 0\\
    \{ F_1(A),F_2(A) \} = & 0\\
    \{ E(f),F(A) \} = & \kappa F(f).
  \end{array}
\end{equation}
The kinematics of General Relativity is then governed by the Gauss-, diffeomorphism- and the scalar constraint $G_i$, $V_a$, $C$, which read in these variables:
\begin{equation}
  \begin{array}{rl}
    G_i = & D_a E^a_i \\
    V_a = & F^i_{ab} E^b_i \\
    C = & \biggl( F^i_{ab} +(\beta^2+1)\epsilon^i_{mn} K^m_a K^n_b \biggr) \frac{\epsilon^{kl}_i E^a_k E^b_l}{\sqrt{det(q)}}.
  \end{array}
\end{equation}
$F$ denotes the curvature of $A$ in the above equations. The dynamics is then governed by the constrained Hamiltonian:
\begin{equation}
  {\bf H}=\int_\Sigma d^3x \biggl( \Lambda^i G_i + N^a V_a + N C \biggr),
\end{equation}
where $\Lambda$ denotes an $SU(2)$-gauge, $N^a$ the shift vector field and $N$ the lapse function.

\subsection{Stratified Diffeomorphisms}

We will need the notion of a stratified diffeomorphism, because a subset of the stratified diffeomorphisms will be investigated later. We will follow the definition given by Fleischhack\cite{fleischhack}, who himself is closely following Hardt\cite{hardt}.
\begin{defi}
  Let $\mathbb X$ be a manifold of differentiability category $p$, and $U$ be a subset of $\mathbb X$.
  \\
  Then: 
  \\
  $\mathcal M$ is called a {\bf stratification} of $U$, iff it is a locally finite, disjoint decomposition of of $\mathbb X$ into connected embedded $C^p$  manifolds $\mathbb X_i$ of $\mathbb X$, such that:
  \begin{equation}
    \mathbb X_i \cap \partial \mathbb X_j \neq \emptyset \Rightarrow \mathbb X_i \subseteq \partial \mathbb X_j \textrm{ and : dim} \mathbb X_i < \textrm{dim} \mathbb X_j.
  \end{equation}
  \\
  The elements $\mathbb X_i$ of the decomposition are called {\bf strata}.
  \\
  $\mathcal M$ is called a stratification of $U$, iff $U$ is the union of some elements of $\mathcal M$.
\end{defi}
We will use this definition of stratification to define stratified maps and particularly stratified diffeomorphisms:
\begin{defi}
  Let $f$ be a continuous map from a $C^p$-manifold $\mathbb X$ to a $C^p$-manifold $\mathbb Y$. The map $f$ is called
  \begin{itemize}
    \item a {\bf stratified map}, iff there is a pair of stratifications $\mathcal M,\mathcal N$ of $\mathbb X$ resp. $\mathbb Y$, and for each stratum $\mathbb X_i$ there exists an open neighbourhood $U_i$ and a $C^p$ map $f_i: \mathbb X_i \subseteq U_i \rightarrow \mathbb X$ with:
    $\overline{\mathbb X_i}\subseteq U_i$, $f_i|_{\mathbb X_i}=f|_{\mathbb X_i}$, $f_i(\mathbb X_i)\in \mathcal N$ and rank$f|_{\mathbb X_i}=$dim$f(\mathbb X_i)$.
    \item a {\bf stratified diffeomorphism}, iff $f|_{\mathbb X_i}$ is injective and and the restriction of each $f_i$ to the respective $U_i$ is a $C^p$-diffeomorphism.
  \end{itemize}
\end{defi}

\subsection{Loop Quantum Gravity}

The objective of quantization is to find a set of variables, that separates the points in the classical phase space and to find a Hilbert-space representation that implements the Poisson-algebra amongst them as $i$-times commutators of the respective operators.
\\
The basic choice of variables for loop quantum gravity on a $3-$manifold $\Sigma$ are holonomies $h_e(A)$ of an $SU(2)$ connection along edges of piecewise analytic paths $e$ in $\Sigma$ and fluxes $P_S(E)$ through analytic surfaces $S$ in $\Sigma$ (which can be defined without reference to any background structure and which separate the points of the classical phase space):
\begin{equation}
  \begin{array}{rl}
    h_e(A):= & P \exp \biggl( \int dt A^i_a(e(t)) \dot{e}^a(t) \tau_i \biggr), \\
    P_S(E):= & \int d^2 y E^a_i(S(y))n_{S,a}(S(y)) \tau^i,
  \end{array}
\end{equation}
where $P$ denotes path ordering of the exponential, $\dot{e}^a$ denotes the tangent vector along $e$ in parametrization $t$ and $n_{S,a}$ denotes the co-normal to $s$ in the parametrization $y$ of $S$. The Poisson-brackets among the holonomies and the fluxes can be calculated by regularizing the edges and surfaces in three dimensions and then taking the limit of a family of functions, that converges exactly to the holonomy along the particular edge and the flux through the particular surface. The result for the Poisson-bracket of a holonomy with a flux can be summarized as follows: first split the edge into pieces, which are either completely in the surface completely outside the surface, such the splits outside have at most a beginning or an endpoint on the surface. For those edges, we have:
\begin{equation}
  \{ h_e(A), P_{S,j}(E) \} = - \frac{l(S,e)}{2} \biggl( h_e(A) \tau_j \,e\textrm{outgoing}, -\tau_j h_e(A)\, e \textrm{incoming}\biggr),
\end{equation}
where $l(S,e)$ gives $1$ for $e$ above $S$, $-1$ for $e$ beneath $S$ and vanishes for completely inside or outside laying edges.
\\
Moreover, the holonomies Poisson-commute, hence, we can build cylindrical functions, which are defined to be functions, that can be written as the composition of a bounded complex-valued function on $SU(2)^n$ and the holonomies along $n$ edges, that intersect at most in vertices. By noticing, that a connection defines a morphism from the path-groupoid into $SU(2)$, we can define a quantum connection to be any such morphism: $A: e \mapsto h_e$. The quantum configuration space $\mathcal C$  is the space of these morphisms and the unique Borel-probability measure $\mu_{AL}$, that is defined by the integration over cylindrical functions by:
\begin{equation}
  \int_{\mathcal C} d\mu_{AL}(A) f(h_{e_1}(A),...,h_{e_n}(A)) = \int_{SU(2)^n} d\mu_H(g_1)...d\mu_H(g_n) f(g_1,...,g_n),
\end{equation}
is called the Ashtekar-Lewandowski measure, arising form the Haar-measure $\mu_H$ on $SU(2)$. The $L^2$-space $L^2(\mathcal C,\mu_{AL})$ is the Hilbert-space, that is used in Loop Quantum Gravity and the cylindrical functions act as multiplication operators thereon. The sup-norm completion of the algebra of cylindrical functions is the algebra of quantum configuration variables $C(\mathcal C)$. One could try to represent the flux operators directly on this Hilbert-space, however they would act as derivative operators and are thus not bounded on the Hilbert-space. Therefore one uses the unitary (Weyl-type) operators, that arise from exponentiating these fluxes. These act as a group $\mathcal W$ of homeomorphisms on the space of groupoid morphisms $\mathcal C$.
\\
Let us now consider an orthonormal basis for this Hilbert-space: First, we know by the Peter-Weyl-theorem, that the matrix elements of the irreducible representations furnish such a basis for $L^2(SU(2),\mu_H)$. Let us now consider graphs $\gamma$, i.e. finite sets of  piecewise analytical edges $e$, that intersect at most in their endpoints, which are the vertices of the graph. Choosing a nontrivial irreducible representation $i_j$ for each edge $e_j$, we see, that there is an orthonormal basis for the Hilbert-space defined by:
\begin{equation}
  \langle T_{\gamma,\vec{i},\vec{m},\vec{n}},A\rangle = \Pi_{e_j\in\gamma}\pi^{i_j}(h_{e_j}(A))_{m_j,n_j}.
\end{equation}
We call the functions $T_{\gamma,\vec{i},\vec{m},\vec{n}}$ gauge-variant spin-network functions and notice, that they are orthogonal, iff they depend on two different graphs.
Hence we see, that there is an overcountable number of graphs, since there is an overcountable number of piecewise analytic edges, thus an overcountable basis for the Hilbert-space. Thus, the Hilbert-space is not separable. However, we still have to factor out the action of the gauge transformations. This is done by associating to each basis element its orbit under these gauge transformations and using the standard technique of an antilinear rigging map $\eta$ to cut down the size of the Hilbert-space. Instead of having  representatives as labels for the basis, only the gauge orbits remain labels of the basis-elements after rigging. 
\\
The main argument of this paper is, that the Hilbert-space of Loop-Quantum Gravity after the physical group of diffeomorphisms is factored out becomes separable. That is, that the orbits of graphs under this group forms a countable set.

\subsection{Statement of the Problem}

We have defined Loop Quantum Gravity as a diffeomorphism-invariant theory, based on holonomies as fundamental variables. A basis for the kinematical Hilbert-space is labelled by gauge-invariant spin-networks depending on holonomies on graphs. Thus, we expect, that the diffeomorphism invariant Hilbert-space has a basis, that depends on spin-networks on diffeomorphism classes of graphs.
\\
This is the point, where the problem arises: consider a vertex in a graph with more than four adjacent edges. Let us consider a chart around this vertex: A diffeomorphism at the vertex can only map three tangent vectors onto three others, thus we can build invariants from the fourth tangent vector on (of the type of relations of length and angles of more than three vectors), that are invariant under diffeomorphisms. Thus, using graphs, the diffeomorphism classes are labelled by these relations, which are continuous sets of real numbers. This means, that the basis of the Hilbert-space contains continuous labels, and there is no known way to escape this continuity. This in turn implies, that the Hilbert-space of loop quantum gravity would be non-separable even after the action of the diffeomorphism group is factored out.
\\
Although this fact is bothersome, this could still be remedied by seeking the solution space of the Hamilton constraint. However, we would like to investigate the following question: Was one too fast as one  based the quantization on loops as fundamental variables and at the same time simply implied  classical diffeomorphism invariance?
\\
Holonomies are not supported by the classical Poisson-bracket after all, hence we would have to regularize them to be able to act with diffeomorphisms at the classical level. The diffeomorphism invariant quantum theory should be build from series of diffeomorphism classes of classical objects. So, we raise the question: Are the diffeomorphism classes of classically regularized cylindrical functions also labelled by such continuous sets? Or, if the answer is no: How can we then obtain a group of transformations (extending the classical diffeomorphisms), such that the orbits under these transformations are exactly series of classically regularized cylindrical functions? (The construction presented in this paper will be given in terms of one-parameter families, containing a limit element at zero. The definitions, that we will introduce later will generally hold for series of objects converging to a limit element. Hence, throughout this paper, we will associate to the one-parameter family of objects series by evaluating the one-parameter family at a strictly decreasing sequence converging to zero.)
\\
We will call such an extension a complete group of gauge transformations, and hold it to be more fundamental than the classical group of gauge transformations, sine a quantum gauge theory is build as a completion of functions depending only on gauge orbits of classical objects.
% \section{Classical and Quantum Gauge Transformations}
  % Classical and Quantum Gauge Transformations

\section{Classical and Quantum Gauge Transformations}

In this section we will first exemplify the enlargement of the classical configuration space to a quantum configuration space, which is usually a larger topological space. Thus, given a group acting as homomorphisms on the classical configuration space, there is usually not an obvious extension of this action to the quantum configuration space. In the case of a gauge theory however, we want to construct a gauge-invariant Hilbert-space, that arises as a Hilbert-space completion of the classical  gauge-invariant functions. This goal in mind gives us the guideline to construct a ''completion'' of the group of classical gauge transformations and an action on the quantum configuration space. This is related (in a sense) to the requirement of completeness, when we define an incompleteness to be a two quantum variables, that are not gauge equivalent, although they arise as the limits of series of classical variables, that are elementwise gauge equivalent.

\subsection{Enlargement of the configuration space}

A classical field theory is usually built upon a Hamiltonian system $\{\Gamma,\{.,.\},H\}$, where the phase space $\Gamma$ often consists of a cotangent bundle over a configuration space $\mathcal C$. The configuration space is usually chosen to be some smoothness class of field configurations on a base manifold, which is chosen due to convenience and has usually little physical significance. These spaces have usually an infinite number of degrees of freedom. The Hilbert-space of the corresponding quantum theory consists of functions of these elementary configuration-degrees of freedom, that are square integrable with respect to some measure. The Hilbert-space completion will now include completions of infinite sums of smooth functions, which are not necessarily smooth, since they only need to be square integrable. The quantum configuration space $\bar{\mathcal C}$ is the spectrum of the commutative algebra of compact multiplication operators on this Hilbert-space, which obviously contains the non-smooth extensions. This is how the quantum enlargement of the classical configuration space of a fields theory comes about.
\\
Let us be more concrete: Under rather general circumstances and most conveniently, one can write the configuration space of a field theory as a space of (in a classical sense) smooth morphisms from some groupoid $\mathcal G$ into some group $\mathbb G$\footnote{Although this is not the usual way to introduce the quantum configuration space of background dependent free Klein-Gordon fields, one can consider a groupoid that is actually the group of modes (e.g. Schwartz functions) together with pointwise addition as composition law, and view the quantum field configurations as a particular completion of a sufficiently large set of nice morphisms from the mode groupoid to $(\mathbb R,+)$. An important difference to Loop Quantum Gravity is, that the group $(\mathbb R,+)$ is not compact and thus the quantum configuration space, which arises as an infinite product of these non compact spaces is not even locally compact.}. This means for all $x \in \mathcal C, g \in \mathcal G$:
\begin{equation}
  x: \mathcal G \rightarrow \mathbb G \,{\textrm{smooth, by:}}\, g \mapsto x(g).
\end{equation}
Each such smooth map defines a configuration variable. The configuration observables are then complex-valued functions of these morphisms $x$. In order to define the commutative $C^*$-algebra of configuration variables, we need a topology to define smoothness. However, by the Gelfand-Naimark theorem, we know that this is a two-way road, namely, that we can define an algebra of continuous functions and a $C^*$-norm thereon, and we get the topology of the spectrum by the weak-operator-topology. Thus we define such an algebra: we consider {\it cylindrical functions}, these are build form functions $f:\mathbb G^n\rightarrow \mathbb C$ (continuous) and a collection $\gamma=(g_1,...,g_n)$ of {\it independent} groupoid elements, then $Cyl$ is a cylindrical function, if there exist $f,\gamma$ s.t.:
\begin{equation}
  Cyl: x \mapsto f(g_1(x),...,g_n(x)).
\end{equation}
The norm for a cylindrical function is the sup-norm, which is clearly a $C^*$-norm. The $C^*$-completion of this algebra can be presented by compact multiplication operators on a Hilbert-space that is build from a {\it uniform measure}: Given a probability measure $\mu_g$ on $\mathbb G$, we can define a measure on $\mathbb G^n$ by the taking the product measure. Now given any collection $\gamma$ of $n$ independent groupoid elements, we associate $\mu^n$ to the functions cylindrical w.r.t. $\gamma$. This such defined measure turns out to be a measure on the spectrum of the configuration variable algebra, if $\mu$ is a probability measure, that is compatible with the groupoid composition. The quantum configuration space is the spectrum of this algebra (in the weak-operator topology). This space contains now elements, that are not in the classical configuration space. 
\\
This quantum enlargement has an important consequence: Given a group of homomorphisms of the classical configuration space, (if it is possible) one needs to have an embedding of the classical configuration space as a dense subset in the quantum configuration space and extend the action by continuity. This could be done by replacing each quantum configuration variable with a one-parameter family of classical variables and consider the continuous extension of the action of the transformation group in the limit in which the quantum observable is attained, i.e. for a transformation $G$ and a quantum configuration variable $v$ with one-parameter approximation $v_\epsilon$:
\begin{equation}
  G v := \lim_{\epsilon \rightarrow 0} G v_\epsilon.
\end{equation}
However, even if such construction can be applied, this will in general not yield the complete group of gauge transformations, as we will see in the next subsection. Note that the limit structure is two-fold in the case of Loop Quantum Gravity: first, there is the Hilbert space completion and second by using loop variables, one uses a classically well behaved, however singular smearing upon which quantization is based, which also just exists as a limit. However, the techniques presented in this paper are independent of the origin of the limit structure and apply whenever a configuration space is enlarged in a certain limit.

\subsection{Extension of Classical Gauge Transformations to the Quantum Configuration Space}

Up to now, we only considered classical field theories without constraints. However, almost all field theories in physics are gauge theories, thus we need to consider classical systems of the kind $\{\Gamma,\{.,.\},H,\{\chi_i\}_{i\in \mathcal I}\}$, where $\{\chi_i\}_{i\in \mathcal I}$ is a set of constraints. This set of constraints generates a group $T$ of gauge transformations $\tau$ on the classical configuration space. The gauge-invariant configuration variables are precisely those elements $f$ of the classical configuration algebra, that satisfy:
\begin{equation}
  f(x) =  f(\tau x).
\end{equation}
Thus, we are interested in configuration variables, that are exactly constant along the orbits generated by $T$. Furthermore, the physical Hilbert-space (i.e. the gauge-invariant Hilbert-space) shall be a Hilbert-space completion of this algebra of gauge-invariant configuration variables. This is no restriction of generality, since we are dealing with a commutative $C^*$-algebra, and we know, that any representation of a commutative $C^*$-algebra is a direct sum of GNS-representations. 
\\
This requirement however, that the physical Hilbert-space is a completion of gauge-invariant configuration variables only leads us to a requirement for the group of quantum gauge transformations: The quantum gauge transformations have to factor out all those quantum configuration variables that are not in the completion of (or do not arise as a limit of) the gauge-invariant configuration variables. Let us consider a triple $(\mathcal H,\pi,U)$, where $\pi$ is a representation of the classical configuration algebra on a Hilbert space $\mathcal H$ by multiplication operators and $U$ is a unitary representation of $T$, then we can define an incompleteness by:
\begin{defi}
  An {\bf incompleteness} is a pair of elements $h_1,h_2 \in \mathcal H$, that are the limits of two sequences, which are elementwise gauge-equivalent (at the classical level), but there exists no element in $\tau$ in $T$, such that $h_1=U(\tau) h_2$.
\end{defi}
Thus, we want is that the group of gauge transformations contains enough elements, such that there are no incompletenesses left. This is a nontrivial issue: Consider for example electrodynamics: The gauge-invariant configuration variables depend only on the transversal degrees of freedom. However, if we first construct a one-particle Hilbert-space containing also longitudinal degrees of freedom, then the completion of the one-particle Hilbert-space contains distributional extensions of longitudinal degrees of freedom, that can not be gauged away with by any classical gauge transformation. Thus, if we use cylindrical functions based on a finite number of modes in this one-particle Hilbert-space and if we solve the constraint by identifying only those quantum configurations, that are related by classical (and thus smooth) gauge transformations, we will find moduli corresponding to these distributional extensions, that cannot be gauge away. However, using for example quantum Faddeev-Popov-fields (as it is standard in background-dependent gauge-field theory), which contain exactly those distributional extensions that the classical gauge transformations could not factor out, we can construct a gauge-invariant Hilbert-space, that is free of these incompletenesses. ( For an example that mirrors the situation of loop quantum gravity better, see the appendix for a similar discussion.)
\\
To put our requirement into a picture, we want that for any pair $v^1_\epsilon, v^2_\epsilon$ of sequences or one-parameter families that yields a pair quantum configuration variables $v^1,v^2$, that if for every $\epsilon > 0$ there exists a classical $\tau \in T$ in the first line, then there exist a quantum $\tau_q$ in the last, such that the following diagram commutes:
\begin{equation}
  \begin{array}{rrcl}
    \tau :& v^1_\epsilon  & \longrightarrow & v^2_\epsilon\\
    \epsilon \rightarrow 0& \downarrow &                   & \downarrow \\
    \tau_q: & v^1 & \longrightarrow & v^2.
  \end{array}
\end{equation}
\begin{defi}
  A $T_q$ (containing all $\tau_q$) is complete on $(\mathcal H,\pi)$, iff there exists a regularization scheme s.t. the above diagram commutes.
\end{defi}
Since we developed our notion of completeness (resp. incompleteness) purely the physical principle, that the physical Hilbert-space of a gauge-theory should be a completion of gauge-invariant quantities only, which we believe to be very reasonable, we will impose it for the case of Loop Quantum Gravity.
% \section{Extension of the Diffeomorphism Group to the Quantum Configuration space}
  % Extension of the Diffeomorphism Group to the Quantum Configuration Space

\section{Extension of the Diffeomorphism Group to the Quantum Configuration space}

\subsection{Regularized Holonomies}

The configuration variables of Loop Quantum Gravity are functions of holonomies, as it was described in section 2. However, what we did not consider in that section is, that holonomies are smeared distributionally and hence the classical Poisson-bracket is not supported unless one regularizes the expression for the holonomies. Recalling, that a holonomy is the parallel-transport element along an edge $e$ and thus the path ordered integral of the exponential of a connection along $e$ : $h_e(A):=P_t\{ \exp(\int_0^1 dt \dot{e}(t)_a A^a_i(e(t))\tau^i)\}$, one realizes, that 
\begin{equation}
  h^\epsilon_e(A)=P_t \biggl\{ \exp\biggl(\int_0^1 \int_{reg_\epsilon(e)} ds d^2\sigma \delta^\epsilon_0(\sigma) \delta^\epsilon_t(s) \dot{p}^a(s,\sigma)A^i_a(p(s,\sigma))\tau_i\biggr)\biggr\}
\end{equation}
is a regularization of a holonomy, which is converging to the particular holonomy in the limit $\epsilon \rightarrow 0$ (compare e.g. \cite{thiemann}).
Here, $p(t,\sigma)$ is a 2-parameter family of mutually nonintersecting paths, that are coincide with $e$ for $\sigma \rightarrow (0,0)$ and the $\delta^\epsilon$ are regularizations of the Dirac delta over the region of integration. The region of integration is an open set $reg_\epsilon(e)$, that is centered around $e$, with the properties, that we will define to be a regularization of $e$. This needs to be made more precise:
\begin{defi}
  Let $e$ be an edge, then a continuous one-parameter family of open sets $\sigma^\epsilon_e$ is called a regulator of $e$, iff 
  \begin{enumerate}
    \item $\partial e$ is in $\partial \sigma^\epsilon_e$ for all $\epsilon > 0$
    \item the interior of $e$ is in $\sigma^\epsilon_e$ for all $\epsilon > 0$
    \item for all $x$ outside $e$, there exist $t> 0$, s.t. $x$ is outside $\sigma^\epsilon_e$ $\forall t > \epsilon$.
  \end{enumerate}
\end{defi}
\begin{defi}
  Let $\sigma^\epsilon_e$ be a regulator of $e$, then a continuous one-parameter family $i_\epsilon (\sigma_e)$ is an approximation of $e$, iff:
  \begin{enumerate}
    \item for all $\epsilon > 0$: $\partial i_\epsilon (\sigma) \cup i_\epsilon (\sigma)$ is a subset of $\sigma^\epsilon_e$
    \item for all $x$ in the interior of $e$, there exists $t > 0$ s.t. $x \in i_\epsilon (\sigma) \forall t > \epsilon$.
  \end{enumerate}
\end{defi}
Using these two definitions, we can define a regularization of an edge as:
\begin{defi}
  A {\bf regularization} of an edge $e$ is a pair $\sigma_e,i(\sigma)$ consisting of a regulator $\sigma_e$ and an approximation $i(\sigma)$.
\end{defi}
\bigskip

\epsfig{file=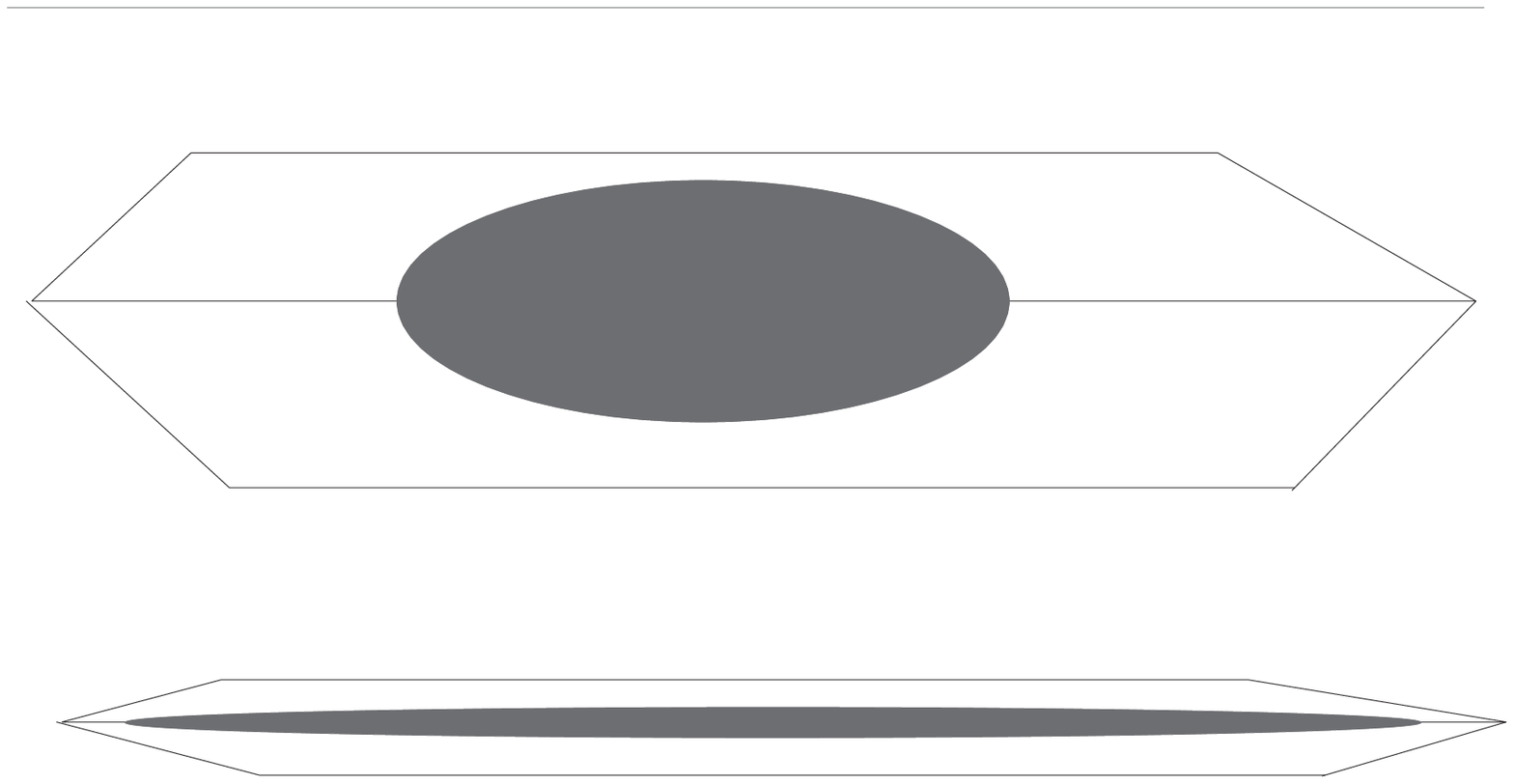, width=12 cm} 
\\
figure 1: {\it{Regularisation of an edge: The first line depicts the unregularized edge, the second line depicts an edge with $\epsilon =1$ and the last line depicts an edge with $\epsilon < 1$. The region $i_{\epsilon}(\sigma)$ is depicted in grey and can be seen to ''shrink down to the edge'' and ''spread all over the edge towards the vertices at the endpoints''.}}
\bigskip

Now, we impose, that the region of integration is given by a regularization of $e$: $reg_\epsilon(e):=i_\epsilon(\sigma_e)$. This is entirely consistent, since the integral over any region and any classical connection depends just on the interior, which is exactly approximated. To avoid confusion, we note, that the regularizations of the Dirac-$\delta$ have to be compatible with the regularization, i.e. the normalization is taken w.r.t. the respective region. Moreover, we notice that we did not introduce or require a background structure for our definition of $reg_\epsilon(e)$, which is thus background independent.

\subsection{Regularized Cylindrical Functions}

The configuration variables of Loop Quantum Gravity are (norm completions of series of) cylindrical functions. Thus, we need to define a regularized and hence the classical version of a cylindrical function. In the previous subsection we regularized holonomies along edges, so they would be supported on by the classical Poisson-bracket. Since cylindrical functions depend on graphs, we will seek a definition of a regularization of a graph, consisting of compatibly regularized edges, and define a classical cylindrical function as a function depending on the regularized holonomies of the regularized graph.
\\
Let us first consider a graph, this is a collection $(E,V)$, where $E$ is a set of piecewise analytical edges, that intersect at most in the vertices $V$, which are the boundaries of the edges. We can keep the vertices for now fixed, since cylindrical functions just depend on edges. The edges however need regularization and we want it to be in such a way, that two different edges do not test the connection on the same points. This is due to the fact, that we want to investigate diffeomorphism classes of regularizations and hence we have to require, that taking any vertex out of the graph, we have to ensure, that the regularization has the same homotopy class as the approximated graph. There are graphs, that contain subsets edges, that intersect parallely in vertices. Due to the definition of a graph, these edges do just coincide at the vertex. However, in general it will be impossible to find two open regions of the edges, whose boundaries just coincide in the vertex. Thus, we have to allow small 1-parameter regions converging to the vertices, in which the boundaries of the regularizations of the parallel edges are allowed to coincide. A workable definition of a vertex-regularization of $v$ is a continuous one parameter family $v^\epsilon$ of neighbourhoods of $v$, such that for every $x \neq v$ there exists a $t > 0$ such that $x$ is outside $v^\epsilon \forall t > \epsilon$.
\epsfig{file=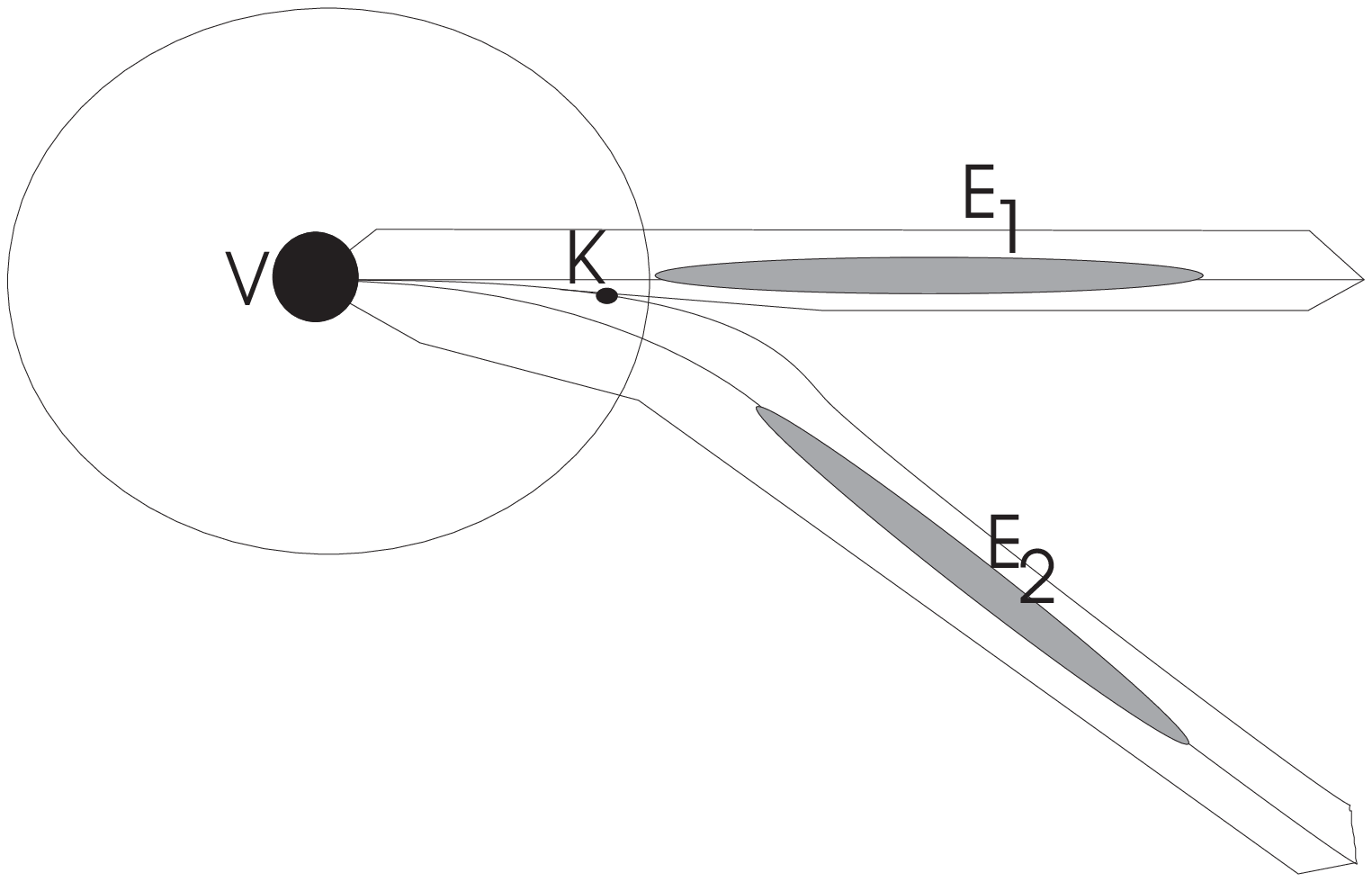, width=12 cm} 
\bigskip

figure 2: {\it{Regularisation of a graph consisting of two edges: Here we depict the problematic situation, where two edges $E_1,E_2$ arrive with the same tangent vector at the vertex $V$. However, due to analyticity, there is no finite region, where $E_1$ and $E_2$ coincide. In this case, we allow, that the boundaries of the regions $\sigma_{E_i}$ partially coincide inside a regularization of the vertex (from $K$ to $V$), which is indicated by the thin lined circle around the vertex $V$.}}
\bigskip

Thus, we arrive at a definition of a regularization of a graph:
\begin{defi}
  A {\bf regularization of a graph} $(E,V)$ is a set $(R,V)$, where for each edge $e\in E$ there is a regularization $i(\sigma_e)$, such that $\forall \epsilon > 0$:
  \begin{enumerate}
    \item for each vertex $v \in V$ and all adjacent edges $e(v)$: $v\in \partial \sigma_e$
    \item for any two non-parallel edges $e_1,e_2$, the intersection $(\partial \sigma_{e_1}\cup \sigma_{e_1})\cap (\partial \sigma_{e_2} \cup \sigma_{e_2})$ is either a common vertex or empty.
    \item for a set of edges intersecting parallel at a vertex $v$ there exists a vertex regularization $v^\epsilon$, such that the $\sigma_e$ of the parallel edges intersect just inside $v^\epsilon$.
    \item the $i(\sigma_e)$ for any two edges do not coincide.
  \end{enumerate}
\end{defi}
\begin{lem}
  Given a graph $\gamma=(E,V)$ in a manifold, there exists a regularization $\gamma_r=(R,V)$.
\end{lem}
proof: The requirements for a regularization can be fulfilled in $\mathbb R^d$. Since they are local requirements, one can fulfill them chart by chart and patch them together. $\square$
\\
Now, the obvious definition for a regularized cylindrical function is:
\begin{defi}
  Given a cylindrical function a representative $f\circ h(\gamma)$, we call $f \circ h_r(\gamma_r)$ a regularization of a cylindrical function, iff $h_r$ is the expression of a regularized holonomy and $\gamma_r$ is a regularization of $\gamma$.
\end{defi}
Notice, that our definition of a regularized cylindrical function does not imply a particular regularisation, but leaves way to take any representative of the equivalence class of regularizations, that result in the same cylindrical function in the limit $\epsilon \rightarrow 0$.
\\
Let us now investigate the properties of diffeomorphism-classes of regularized cylindrical functions.

\section{Nicely Stratified Diffeomorphisms}

Let us consider the regularizations for a given cylindrical function depending on a graph $\gamma$ and investigate its transformation properties under a special class of $\gamma$-nicely stratified diffemorphisms. These are defined as follows:
\begin{defi}
  A stratification $\mathcal M$ in the $d$-dimensional manifold in which $\gamma$ is embedded is called {\bf $\gamma$-nice}, iff 
  \begin{enumerate}
    \item the interior of each edge is completely contained in a $d$-dimensional stratum
    \item all of $\gamma$ (including the vertices) lies in the stratification (i.e. the vertices are allowed to lie in less than $d$-dimensional strata).
  \end{enumerate}
\end{defi}
It follows immediately:
\begin{cor}
  For each stratification $\mathcal M$ and each graph, having no parallel intersections with a less than $d$-dimensional stratum, there exists a decomposition $\gamma^\prime$ of $\gamma$, such that $\mathcal M$ is $\gamma^\prime$-nice. \label{nice-cor}
\end{cor}
Using the notion of a $\gamma$-nicely stratified diffemorphisms we can formulate the lemma:
\begin{lem}
  For the base manifold $\Sigma$, let there be given:
  \begin{itemize}
    \item $U\subset \Sigma$ smooth
    \item $\gamma$ smoothly embedded graph in $U$
    \item $\mathcal M$ a $\gamma$-nice stratification of $U$
    \item $\epsilon > 0$
    \item $\phi_\mu$ a $\mathcal M$ smooth stratified diffeomorphism
  \end{itemize}
  then there exists a regularization $\gamma_r$ of $\gamma$ and a smooth diffeomorphism $\phi$, such that $\phi_\mu(\gamma_r^\epsilon)=\phi(\gamma_r^\epsilon)$ (where $\gamma_r^\epsilon$ denotes the set of points of the union of all $i_\epsilon(\sigma_e)$).
\end{lem}
proof: the interior of the degrees can be mapped by a diffeomorphism. We need to consider the neighbourhood of the vertices (let us first consider vertices with just non-parallel adjacent edges): thus for a given vertex fix a chart around each edge take a tube of coordinate diameter $\epsilon$. Exclude from each tube those parts, that are either not in the stratum or whose coordinate distance to another edge is less than twice the distance the the edge corresponding to the tube. Take the interior of these tubes as $\sigma^\epsilon_e$. Then take an interior open region of $\sigma^\epsilon_e$ with analytical boundary of coordinate distance at least $\epsilon / 3$ and at most $2 \epsilon / 3$ from the boundary of $\sigma^\epsilon_e$ and denote it by $i_\epsilon(\sigma^\epsilon_e)$. Note, that the union of all $i_\epsilon(\sigma^\epsilon_e)$ furnishes a regularization of $\gamma$. Using this regularization, we have no problem to finding a smooth diffeomorphism $\phi$, that has the same effect as $\phi_\mu$ on the respective $i_\epsilon(\sigma^\epsilon_e)$ for fixed $\epsilon > 0$, since the non-smooth regions of $\phi_\mu$ are not in $i_\epsilon \cup \partial i_\epsilon$. 
\\
For vertices with parallel intersecting edges choose take a coordinate ball of size $\epsilon / 2$ centered around the vertex and these as $v^\epsilon$. Take tubes around the edges and cut off the same regions as above, except those points, whose coordinate distance is less than $\epsilon / 4$. Now choose an interior $i_\epsilon(\sigma)$, that has the same properties as above except, such that it is outside $v^\epsilon$, but such, that there is a point with coordinate distance at most $3 \epsilon / 4$. Now, the same argument as above applies for the diffeomorphism $\phi$.
$\square$
\\
Using the regularization constructed in the proof above, we can regularize cylindrical functions and thus we have the corollary:
\begin{cor}
  For any cylindrical function $f$ (w.r.t. a graph $\gamma$), any $\gamma$-nice stratified diffemorphism $\phi_\mu$, any $\epsilon > 0$ there exists a smooth diffeomorphism $\phi$, such that $\phi_\mu^* f = \phi^* f$.
\end{cor}
Thus, we when we try to construct the complete group group of gauge transformations, and using regularizations as described above, we can not construct diffeomorphism classes of cylindrical functions that are not invariant under stratified diffeomorphisms for any finite value of the regulator. Moreover, we see, that the piecewise analytical category of edges is preserved by the action of the stratified analytical diffeomorphism group, since any edge is finite and the stratified diffeomorphisms are analytical except in a finite number of points along a finite edge.
Thus, forming a series from the one-parameter family by evaluating it at a strictly decreasing sequence converging to zero, we conclude with the corollary:
\begin{cor}
  The complete gauge-group of diffeomorphisms contains at least the nicely stratified diffeomorphisms.
\end{cor}
 Notice however, that we can not change the topology of the graph at a finite value of the regulator, since the vertices are fixed as boundaries of the $\sigma_e$.
\\
Let us conclude this section by remarking, that our physical principle, namely, that the physical Hilbert-space should consist of completions of gauge-orbits and not of orbits of completions, is in fact realized and formulated in a background independent way. Moreover, we follow exactly those prescriptions, that where used in the derivation of the Poisson-algebra underlying loop quantum gravity (compare e.g. \cite{thiemann} or for a treatment of the fluxes \cite{acz}). Furthermore, it is important to notice, that our approach does not compromise the existence of holonomies along piecewise analytical edges as fundamental degrees of freedom, but rather puts the very procedure that lets us implement holonomies as fundamental operators to the forefront.
% \section{Action of the Quantum Diffemorphism-group}
  % Action of the Annomaly-free Diffeomorphism Group

\section{Action of the Quantum Diffemorphism-group}

\subsection{Definition of the system}

When we defined Loop Quantum Gravity (in section 2), we left two important ingredients out: For one we did not specify the action of the Hamilton transform, and second we did not specify the action of the diffeomorphism group. The first is a topic of an investigation that seems to be far from complete, the second however, we hope to give a satisfying answer in this section:
\\
Let $C(\bar{\mathbb X})$ denote the sup-norm completion of the algebra of cylindrical functions. Let $\mathcal S$ be the set of stratified analytical surfaces and let $\mathcal W$ be the group of Weyl-operators corresponding to the elements of $\mathcal S$ as described in section 2 and let $\mathfrak A$ denote the norm-completion of the corresponding Weyl-algebra. 
\\
Following the discussion of the last section, we would like to include the nicely stratified diffeomorphisms into the group of our gauge transformations, however the notion of nicely stratified depends on the particular cylindrical function. In the quantum theory however, we have the following lemma:
\begin{lem}
  Given any stratification $\mathcal M$, any $\mathcal M$ stratified diffeomorphism $\phi_\mu$ and any graph $\gamma$, there is a decomposition $\gamma^\prime$ of $\gamma$, such that there is a $\gamma^\prime$-nice stratification $\mathcal M^\prime$ and $\mathcal M^\prime$-stratified diffeomorphism $\phi^\prime_\mu$ with $\phi_\mu(\gamma^\prime)=\phi^\prime_\mu(\gamma^\prime)$.
\end{lem}
proof: Since stratifications are locally finite, an intersection of a less than $d$-dimensional stratum with any edge decomposes into vertices or edges. Now build a stratification such that the extended (edge)-intersections of any less than $d$-dimensional stratum with any edge are blown up into a small enough new d-dimensional stratum by choosing a local chart around the intersection and taking a small enough coordinate tube. Decompose the edges of the graph accordingly. $\square$
\\
Thus, we include the stratified smoothnessclass of diffeomorphisms into the group of gauge-transformations, that we will implement in the quantum theory, which we denote from now on by $\mathcal D$. Their action on a cylindrical function shall be denoted by $\alpha_\phi$ for any $\phi \in \mathcal D$. We will in particular impose that the gauge group of Loop Quantum Gravity contains stratified analytical Homeomorphisms, i.e. Homeomorphisms that arise as a patching together of strata with analytical diffeomorphisms between them.

\subsection{Countable Diffeomorphism Invariant Basis}

Our line of reasoning for the construction of a separable Hilbert-space will be similar to Fairbairn and Rovelli \cite{fair-rov}: We will show, that for any two graphs in the same knot-class there is a stratified analytical diffeomorphism, that maps one onto the other graph, which we will show by explicit construction. Thus, since we identified the stratified diffeomorphisms with the physical group of gauge transformations, we can take the knot-classes of graphs together with the spin-labels as a basis for the Hilbert-space, and conclude that this Hilbert-space is separable, since the basis is countable.
\\
Let there be graphs $\gamma,\gamma^\prime$ that are in the same knot-class. We know from the introduction, that they are in general not mapped onto each other if they have vertices of valence 4 or more. Let us now find a stratification $\mathcal M$ and a stratified diffeomorphism $\phi_\mu$, such that $\phi_\mu(\gamma)=\gamma^\prime$:
\\
Construct two stratifications based on the following construction, one each for $\gamma$ and $\gamma^\prime$: Around each vertex choose a chart, such that $v$ lies at the origin and choose a coordinate ball small enough, such that the closures of the coordinate balls around all vertices have empty intersection. Around each edge $e$ choose a chart such that $e$ lies on the $z$-axis between $z=0$ and $z=1$. Choose an $\epsilon$ small enough for each $e$, such that the closures of the coordinate tubes around each edge have empty intersection outside the balls around the vertices. Then take off the coordinate tubes all those points, whose coordinate distance in the vertex-chart to a different edge are less than twice the coordinate distance to the edge in the respective tube. 
\\
Now take a smooth function $f(z)$, such that the regions and of coordinate distance less than $f(z)$ in each edge chart are all contained in the cut of tube. Call these smaller regions in the tubes Gaussian shaped tubes. Moreover, around each vertex, we take a small region, such that these regions are mutually disjoint. Take the complement of the edge-regularizations, their boundaries and the vertices in these regions. This complement can then be stratified, such that each stratum is at most adjacent to one edge-regularization. A consequence of this is, that we have to refine the stratification of the edge boundaries to obtain a stratification of the entire region containing the graph, such that this refinement is included in our stratification. 
Using these Gaussian shaped tubes, their boundaries, the vertices and a stratification of the complement together with the refinement of the stratification of the boundaries, we have a stratification of a region, that contains the respective graph.
\\
Now, the key observation is, that there are smooth diffeomorphisms mapping the Gaussian shaped tubes, their boundaries, the vertices and the complement of the stratification corresponding to one graph onto the other graph and mapping the edges in the Gaussian shaped tube exactly onto each other. These can be chosen to be stratified diffeomorphisms, iff all charts (from which the stratification was constructed) are of a sufficient differentiability class and then if and only if the two graphs are in the same knot class. Let us summarize the result of this construction in the following lemma:
\begin{lem}
  For any two graphs $\gamma,\gamma^\prime$ in the same knot class, there exists a $\gamma$-$\gamma^\prime$-nicely stratified diffeomorphism (of the differentiability class of the atlas of the underlying manifold), that maps $\gamma$ onto $\gamma^\prime$.
\end{lem}
proof: Using the construction above, we can use the identity maps between the respective edge charts to map the Gaussian-shaped regions and their boundaries according to the differentiability class of the atlas. Then there is a continuous extension of these diffeomorphisms to the vertices and the rest, which can each be chosen according to the differentiability class of the atlas. $\square$
\\
Using the definition of a complete group of quantum gauge transformations, this has an immediate corollary:
\begin{cor}
  The complete diffeomorphism-orbits of a cylindrical function depends only on the knot-class of the underlying graph.
\end{cor}
proof: The stratified-diffeomorphism rigging map maps a spin network $T_\gamma$, depending on a graph $\gamma$ into a linear functional that just depends on the knot-class of $\gamma$. Completeness of the spin-network basis and orthogonality of spin networks depending on different graphs in the diffeomorphism-variant inner product gives the proof. $\square$
\\
Form knot theory, we know, that the knot classes are countable. Since the spin labels also form a countable set, we conclude just as Fairbairn and Rovelli \cite{fair-rov} (consulting the references to knot-theory in their paper):
\begin{lem}
  The complete diffeomorphism-invariant Hilbert-space of loop quantum gravity is separable.
\end{lem}
proof: there is a countable basis formed by the knot-classes of spin-network functions. $\square$
Note, that we could have anticipated this result by trying to build a basis from series of diffeomorphism classes spin-network functions depending on regularized graphs by using the regularization prescription from the previous section.

\subsection{Note on Completely Covariant Operators}

After we identified the nicely stratified diffeomorphisms as the complete group of diffeomorphism transformations, that are to be implemented into the quantum theory, we should investigate whether the operators of quantum geometry transform covariantly under nicely stratified diffeomorphisms, meaning for each stratified diffeomorphism $\phi_\mu$ there exists a unitary operator $U_\phi$ (satisfying the  multiplication  $U_\phi U_\psi = U_{\phi \circ \psi}$ and inversion $U^*_{\phi}=U_{\phi^{-1}}$ of the associated diffeomorphisms):
\begin{equation}
  Geom(\phi_\mu(Obj))=U_\phi Geom(Obj) U^*_{\phi}.
\end{equation}
The geometrical Operators of interest are the area and the volume operator. Proceeding analogous to Fairbairn and Rovelli \cite{fair-rov}, we will see, that the operators, that were identified as covariant under their class of extended diffeomorphisms is indeed covariant under stratified diffeomorphisms:
\\
The key observation here is, that stratified diffeomorphisms do not change the topology of geometrical objects since they are a special class of homomorphisms, and hence do not change topological relations among objects. They do however change geometrical relations, since stratified diffeomorphisms are allowed to have ''kinks''. Thus, we call geometrical operators completely covariant, iff the expectation value of the geometrical quantity (of an object $O$)  in a certain spin-network-state (depending on a graph $\gamma$) does not depend on the differential relations between $\gamma$ and $O$, but just on the topological relations. 
\\
From this discussion, it is obvious, that the FLR-area operator \cite{lehner} and the RS-volume-operator\cite{rov-smol}, which both do only depend on the topological relations of the areas resp. volumes with the considered graph are indeed covariant under the group of stratified diffeomorphisms. Thus, there exists a version of quantum geometry, that is covariant w.r.t. the complete group of diffeomorphism transformations, allowing us to give the group of stratified diffeomorphisms a physical interpretation.
\\
Giesel and Thiemann pointed out in \cite{giesel}, that the RS-volume operator would be inconsistent with the dynamics of the standard version of the Hamilton constraint. Our argument presented here however is entirely kinematical, hence stated in a domain in which the RS-volume operator can be applied consistently. Moreover, the question of actually determining the spectrum of the volume operator is (at least for now) out of the scope of experimental physics, and thus any  Hermitian operator that can be interpreted as a volume in a classical limit has an interpretation in its own right. 
\\
However, since at present only the kinematics of Loop Quantum Gravity is well understood, we would like to caution, that there is no known kinematical reason to rule the Rovelli-Smolin volume operator out as inconsistent. Thus, for kinematical observables, such as areas and volumes, we are free to use any set of operators, that has the transformation properties, that we desire. From a dynamical point of view, it could turn out, that once a completely physical dynamics (i.e. a dynamics w.r.t. physical internal clocks) of Loop Quantum Gravity is developed, that the dynamics tells us that these kinematical operators correspond to different physical clocks, very much like in perturbation background-dependent field theory where renormalization group ''links'' different sets of operators for different scales. The conclusion for loop quantum gravity is, that there may exists more than one physical version of quantum geometry, where two inequivalent ones are related by the dynamics of the theory, which is (unfortunately) not yet understood.

\subsection{Note on the Gauss - and Hamilton-constraint}

The natural question to ask, is: ''Why did this very completion problem not occur with the Gauss constraint?'' The short answer is: holonomies are already partially invariant under the gauge-transformation generated by the Gauss-constraint. Let $G:\Sigma \rightarrow SU(2)$ be smooth, hence defining an element of the classical group of gauge transformations, then the associated unitary operator $U_G$ acts on an holonomy with limit points $x_i,x_f\in \Sigma$ by:
\begin{equation}
  U^*_G h U_G = G(x_i) h G(x_f),
\end{equation}
but it does not act on the interior points of $h$. The cylindrical functions on the other hand always depend on a finite number of holonomies, thus the induced action of the finite Gauss transformations depends only on a a finite number of elements, which using the above notation and numbering the particular edges can be denoted by $\{G(x^{(1)}_i),G(x^{(1)}_f),...,G(x^{(n)}_i),G(x^{(n)}_f)\}$. Thus any class of functions, that can attain any value at any finite number of points is large enough to act as the complete quantum Gauss-group, thus not only the smooth, but even the analytical category form a complete Gauss-quantum gauge group.
\\
Since the standard version \cite{qsd} of the Hamilton constraint acts only on vertices, one would expect, that similar arguments to the one above apply. However, the action of finite Hamilton-transformation is not known, thus one is not even in the position to apply the techniques used in this paper. 
\\
Since the dynamics of loop quantum gravity is still an open question, we will take a different approach: How can we define a Hamilton-constraint along similar to the construction in this paper? A possible answer would be to use regularized cylindrical functions $Cyl^{(\epsilon)}$ (depending on regularized holonomies as above) and defining operators $H(N)$ acting on  $Cyl^{(\epsilon)}$, such that 
\begin{itemize}
  \item the operation on $Cyl^{(\epsilon)}$ does not depend on $\epsilon$,
  \item the limit $lim_{\epsilon \rightarrow 0} H(N) Cyl^{(\epsilon)}$ exists for a dense subset of the cylindrical functions and
  \item the kernel is compatible with the classical kernel: $\{H_{class.}(N),Cyl^{(\epsilon)}\}|_{\epsilon \rightarrow 0}=\mathcal{O}(\epsilon) \Leftrightarrow H(N) Cyl^{(\epsilon)}=\mathcal{O}(\epsilon)$.
\end{itemize}
Then define the kernel of the Wheeler-DeWitt operator through the completion of:
\begin{equation}
  \{F\in Cyl : H(N) F := \lim_{\epsilon \rightarrow 0} H(N) F^{(\epsilon)}=0\}.
\end{equation}
The above procedure again first seeks an orbit and then completes the Hilbert-space, thus doing the analogue of the procedure described in this paper in terms of the generators of the gauge group.
The constraint surface constructed under this principle will in general differ from the one that that arises as the kernel of a set of constraint operators, since the constraint is not linear in the momenta. However, Dirac's prescription to quantize the constraints is just ''a substitute'' for the ''silver bullet'' given by quantizeing the constraint surface directly, which is closer related to the procedure proposed here.
% \section{Conclusions}
  % Conclusions

\section{Conclusions and Outlook}

We considered the group of classical diffeomorphisms as part of the gauge group of General Relativity. We studied how Loop Quantum Gravity is constructed from classical General Relativity. Along this way we found a physical argument for considering nicely stratified diffeomorphisms as the group of quantum diffeomorphisms, by considering what effect gauge transformations have to have in a quantum theory. This comes from the physical requirement, that the gauge-invariant Hilbert-space should be a completion of the space of orbits of gauge-variant configuration variables. 
\\
The restriction to analytical diffeomorphisms seemed, as most of the previous literature deemed to be necessary to preserve the category of piecewise analytical edges turned out to be incomplete (in the sense of groups of gauge transformations). The group of stratified analytic diffeomorphisms on the other hand turned out to be complete, while still preserving the category of piecewise analytical edges.
\\
The physical principle of completing gauge-orbits is particularly motivated by applying a GNS construction, where we want to factor out the ideal of gauge-variant variables from the full variable algebra and then preform a GNS-construction with the truly physical gauge-invariant variables algebra. Starting from the requirement, that the action of a quantum gauge group should yield the same quotient (resp. equivalence classes) as if this the gauge fixing had occurred before quantization, we obtain the definition of a complete group of quantum gauge transformations. 
\\
By going to a suitable classically regularized version of Loop Quantum Gravity, we where able to identify the nicely stratified diffeomorphisms as the complete group of quantum diffeomorphisms, as they are needed for Loop Quantum Gravity.
\\
Consequently we studied the action of this group on kinematical Loop Quantum Gravity. It turned out, that the orbits of graphs are then simply the knot-classes of the particular graphs. This addresses immediately the issue of separability of the diffeomorphism-invariant Hilbert-space of Loop Quantum Gravity, because the knot-classes are countable. A diffeomorphism-invariant spin network on the other hand is labelled by a diffeomorphism-class of a graph and spin quantum numbers associated to each edge. The spin quantum numbers are countable and the knot classes are countable, hence the diffeomorphism-invariant spin networks are countable. Since the diffeomorphism invariant spin networks form a basis for the diffeomorphism-invariant Hilbert-space, we have shown, that this Hilbert-space is separable.
\\
Thus, we have constructed a physical argument, that supports the conjecture and investigations by Fairbairn and Rovelli, that the non-separability of the diffeomorphism-invariant Hilbert-space of Loop Quantum Gravity is spurious. Particularly, we contend that following our line of reasoning, it seems reasonable in the kinematical context to use versions of quantum geometry, that are covariant under stratified diffeomorphisms in order to obtain physical operators.
\\
The covariance of Loop Quantum Gravity under stratified diffeomorphisms has to be demanded at the level of the quantum variable algebra. Since the Hilbert-space (after the diffeomorphism group $\mathcal D_{strat.}$ is factored out) turned out to be separable, we argue, that there is indeed an algebra $\mathfrak A_o$ based on ''diffeomorphism-fixed'' spin knot-functions, such that the kinematical variable algebra can be written as a semidirect product of the diffeomorphisms acting on $\mathfrak A_o$. In a forthcoming work\cite{tim}, we will construct $\mathfrak A_o$ directly and study the representation theory of the full variable algebra by demanding $\mathcal D_{strat.}$-covariance. Moreover, using the separability of the kinematical Hilbert-space, new approaches for the embedding of reduced models into the full theory of loop quantum gravity and their induced representations are investigated in another forthcoming work \cite{tim}.
\\
Future work will also include an investigation into the feasibility of the programme described above for the construction of the kernel of the Wheeler-DeWitt operator.
\subsection*{Acknowledgements:} This work was supported by the Deutsche Forschungsgemeinschaft (DFG). I am particularly grateful to Martin Bojowald for a careful reading of the first draft of this paper and discussions with him. I am also grateful to Christian Fleischhack for a stimulating discussion, that lead me to consider this subject.

\newpage
\begin{appendix}

\section{Example: Two Klein-Gordon Fields Constrained to One}

In this section, we want to give a simple example illustrating the importance of the physical argument of first finding the gauge orbits and then constructing the Hilbert-space. Let us therefore consider quantum field theory of two free real Klein-Gordon fields. The classical field equations are stated for twice differentiable modes. Now, we want to arrive at a theory of just one Klein-Gordon field, thus we consider the difference between the two field strength at each point as pure gauge, thus let us consider the constraint set:
\begin{equation}
  \chi(f):=\pi_1(f)-\pi_2(f)
\end{equation}
for any twice differentiable mode $f$ with compact support. The corresponding finite gauge-transformations are labelled by a mode $f$:
\begin{equation}
  T_t(f):= \exp(t\{\chi(f),.\})
\end{equation}
and the product of two is completely analogous to the product of two momentum-Weyl-operators. Let now $\Psi: \mathbb R^n \rightarrow \mathbb C$ be a smooth function of compact support on $\mathbb R^n$ and let $(f_1,...f_n)$ be a set of continuous modes of compact support.  Then we define a cylindrical function in the standard way as being a functional of the field configurations, that can be written as:
\begin{equation}
  \Psi(\phi_1(f_1),...,\phi_1(f_n),\phi_2(f_1),...\phi_2(f_n)).
\end{equation}
(Note, that there could be a motivation for selecting two different sets of smearings, i.e. two different differentiability classes for the modes, just there it is in loop quantum gravity by the very nature of loops as fundamental variables.) 
\\
Now, suppose, we have a Hilbert-space representation, in which two functions depending on different modes are automatically orthogonal (which is not unitarily equivalent to the standard Fock representation), then using the group averaging procedure to factor out the gauge orbits will not yield the expected result, that the gauge-invariant states are labelled by a completion of the functions of a single field $\phi$, but there will also be remnant moduli corresponding to the lack of smoothness of some of the modes $f_i$. We can obtain these by considering for example $\Psi=F(\phi_1(f)-\phi_2(f))$, for some non-differentiable mode. Then the rigging map $\eta[\Psi](const.)=\sum_{f\in C^2} \langle \Psi , const. \rangle$ vanishes contrary to the seeming equivalence.
\\
But, what happens, when we apply the requirement of a complete quantum gauge group? For any continuous function of compact support, there exists a series of smooth function of compact support approximating it point-wise. Then, using the language from above, we call the lack of group elements with continuous, non-differentiable elements an incompleteness of the quantum gauge group. Thus, a complete quantum gauge group 
contains at least group elements corresponding to all continuous modes and thus removes the moduli, that arise from the non-differentiability of the modes used to construct the pre-Hilbert-space.
\\
The situation in loop quantum gravity is completely analogous: We use 1-dimensional smearing for the connection, but we act on it with the group of diffeomorphisms, that is only known to be complete for a suitable differentiability class of 3-dimensionally smeared objects. Moreover, the gauge-variant inner product is such, that two cylindrical functions based on two different graphs are always orthogonal, even if they can not be distinguished in a 3-dimensional regularization.

\end{appendix}

\end{document}